# Analysis of India's Agricultural Ecosystem using Knowledge-based Tantra Framework

Shreekanth M Prabhu, Professor, Department of Computer Science and Engineering, CMR Institute of Technology, Bengaluru

Natarajan Subramanyam, Professor, Department of Computer Science and Engineering, PES University, Bangalore

## ABSTRACT

The information systems have been extremely useful in managing businesses, enterprises, and public institutions such as government departments. But today's challenges are increasingly about managing ecosystems. Ecosystem is a useful paradigm to better understand a variety of domains such as biology, business, industry, agriculture, and society. In this paper, we look at the Indian Agricultural ecosystem. It is a mammoth task to assimilate the information for the whole ecosystem consisting of consumers, producers, workers, traders, transporters, industry, and Government. There are myriad interventions by the state and the central Governments, whose efficacy is difficult to track and the outcomes hard to assess. A policy intervention that helps one part of the ecosystem can harm the other. In addition, sustainability and ecological considerations are also extremely important. In this paper, we make use of the Knowledge-based Tantra Social Information Management Framework to analyze the Indian Agricultural Ecosystem and build related Knowledge Graphs. Our analysis spans descriptive, normative, and transformative viewpoints. Tantra Framework makes use of concepts from Zachman Framework to manage aspects of social information through different perspectives and concepts from Unified Foundational Ontology (UFO) to represent interrelationships between aspects.

**Keywords: Framework, Ecosystem, Information Management, Information Ecosystem, Ontology, Governance, Agriculture, Knowledge Graph**

**Biographical Notes**

**Dr Shreekanth M Prabhu** is currently working as Professor and Head of the Department of Computer Science and Engineering at CMR institute of Technology, Bengaluru, India. He received his M. Tech. degree in Computer Science and Engineering from Indian Institute of Technology, Bombay in 1986. After that he worked with IT Majors such as TCS, IBM and Hewlett Packard for 25 years. He received his PhD in Computer and Information Science Engineering from Visvesvaraya Technological University, Belagavi, India in 2020. His research interests include Frameworks and Models, Social Networks, E-Governance and Linguistics.

**Dr Natarajan Subramanyam** holds PhD from Jawaharlal Nehru Technological University, Hyderabad (JNTUH), India He is currently Professor and Key Resource Person in Computer Science and Engineering, Department of CSE of PES University, Bangalore, India. He is a Life Member of Computer Society of India (CSI), Indian Society of Remote Sensing (ISRS) etc. He is Reviewer of articles published in Elsevier, Springer, and IEEE etc. Prior to taking up teaching and research in 2005 he was a Senior Scientist at National Remote Sensing Centre (NRSC) of Indian Space Research Organization (ISRO) for nearly three decades. During his stint at NRSC he was Deputy Project Director of Large-Scale Mapping (LSM) Project.

# 1. Introduction

India has been a leader in Agriculture with advanced know-how for millennia. Kautilya's treatise Arthashastra [1] which is dated around 230 BCE gives rich insights into agricultural practices in ancient India. Kautilya believed agriculture was the basis for any economy. He had an exhaustive methodology for cultivation, selection of seeds, forecasting weather conditions, and rainfall. The knowledge about agriculture was highly valued and even King was expected to be well-versed with agriculture.

Kautilya believed in the Yogakshema(well-being) of people, in particular of widows and orphans who needed extra care. He supported a mixed economy with a profitable public sector. Even though Agriculture was regulated there was significant freedom to peasants. Only excess profits of businesses were taxed. Other important recommendations of Kautilya include (i) taxation of farmers, where farmers (ii) setting price based on demand and supply; (iii) price support by Government and practice of maintaining buffer stocks by Government ; (iv) Mandis where farmers were supposed to sell produce and certain restrictions on farmers where they can sell the produce to ensure tax collection (v) regimen of maintaining data related to agriculture including proper land records (vi) acknowledgment of state's role in regulating agricultural markets while giving adequate freedom to farmers and businesses. There is also documentation of crop management in treatises such as Krishi Parashara circa 100 CE, Kashyapiya Krishi Sukti circa 800 CE, and Vriksha Ayurveda circa 1000 CE. The common underlying theme being the supportive role of the king to farmers in terms of market facilitation through traders and giving subsidies to the deserving farmers.

Circling back to the present, India's agriculture sector is one of the largest in the world. India is the second-largest producer of rice, wheat, sugarcane, cotton, groundnuts, fruits, and vegetables, and the largest producer of milk. When it comes to pulses India is the largest producer, consumer, and importer [2]. However, the Indian agriculture sector compared to other countries involves far too many people with a low per-capita contribution to GDP. Public investment in agriculture is stagnant. Agriculture wages also have declined. The government's policies in favour of cereals also have resulted in distortions in cropping patterns and led to producer orientation instead of consumer orientation. Farm markets need to be freed from excessive state intervention [3].

A few decades back challenges in the Indian agriculture sector were about increasing production. Now we have issues where over-production causes loss of income to farmers when the environment is conducive for farming, alternating with losses due to drought. There are also several cases where farmers commit suicide due to a variety of reasons. Most often indebtedness to money-lenders is the major cause. To address the concerns of farmers, Governments have come up with a host of schemes. Thus, Governments in power respond with complex and confounding policy interventions, whose efficacy is hard to assess and monitor. Table 1 below gives details on budgetary allocations the Government of India made to different schemes in the years 2018-19 and 2019-20 [4].

The government also provides subsidies related to power and fertilizer, which are not included in Table 1. Over and above this, State Governments have their allocations and schemes. Despite all these allocations, Alagh opines that many schemes have last mile issues and do not reach farmers [5].

The government of India plays an all-pervasive role in the agricultural sector. Above all, Government tightly regulates agricultural markets and exports. Some laws restrict the sale of farmland for use other than farming. Even the reforms by the Government of India where they passed 3 farm laws [6] to liberalize trade in the farm sector have faced protests by farmers and middlemen. The laws are currently in abeyance.

Table 1: Agricultural Allocations (in Rs. Crores)

| Sr. No | SCHEME | Nature of Support | 2018-19 | 2019-20 |
|---|---|---|---|---|
| 1 | Pradhan Mantri Kisan Samman Yojana (PM-KISAN) | Rs 6,000 per year is provided as Minimum Income Support to Farmers | 20,000 | 75,000 |
| 2 | Interest Subsidy for short-term credit to Farmers | Interest subvention (2% on grant of loan and 3% additional on repayment). The loan has INR 3,00,000 limit. | 14,987 | 18,000 |
| 3 | Pradhan Mantri Fasal Bima Yojana | Crop Loss Insurance. Covers yield-losses, prevention of harvesting due to weather, post-harvest losses, and local calamities at individual farm level. | 12,976 | 14,000 |
| 4 | Rastriya Krishi Vikas Yojana | Central Grants to State Governments to invest in agriculture | 3,600 | 3,745 |
| 5 | Pradhan Mantri Krishi Sinchai Yojana | Extend cover of irrigation and increase the efficiency of water usage i.e., more crop per drop. | 2,955 | 3,500 |
| 6 | Market Intervention Scheme and Price Support Scheme | MIS takes care of horticultural products and PSS takes care of the Minimum Support Price (MSP) for 22 commodities | 2,000 | 3,000 |
| 7 | National Mission for Horticulture | Center and State contribute at 60:40 ratio for integrated development of horticulture | 2,100 | 2,226 |
| 8 | National Food Security Mission | Food grains (rice and wheat) are provided at very low rates to nearly 66% of the population. The outlay is in Lakhs of Crores). | 1,510 | 2,000 |
| 9 | Pradhan Mantri Annadata Aay SanraksHan Yojana | PM-AASHA is an umbrella scheme that combines 3 schemes: Price-Support (PPS), Price-deficiency Payment Scheme (PDPS), and Pilot Private Procurement and Stockist Scheme (PPPS). | 1,400 | 1,500 |
| 10 | Integrated Scheme on Agricultural Marketing | This includes Marketing infrastructure including storage, support for integration of value chain, and electronic marketing. | 500 | 600 |
|  | Total |  | 67,800 | 1,30,485 |

Despite the heavy intervention of the Government agricultural sector continues to face some calamity or conflict. This situation calls for advanced information systems that can provide a comprehensive analysis of the Indian Agricultural Sector. This is best done using an ecosystem paradigm. The analysis of the Indian Agricultural System needs to happen with three different viewpoints: descriptive, normative, and transformative. When we use a descriptive viewpoint, the objective is to describe ecosystem information in a unified manner as consistently as possible. With the normative viewpoint, the objective is to design the information system so that it can help achieve a set of goals for the ecosystem. With a transformative viewpoint, the idea is to take the ecosystem to the next stage of evolution. Here ecosystem participants compete as well as cooperate. They act as agents who may self-organize coevolve and adapt to the environment and collectively be part of emergent phenomena.

Johnson [7] emphasizes that the agricultural sector needs a multi-disciplinary approach that requires excellence in multiple disciplines instead of an inter-disciplinary approach that concerns the interface between multiple disciplines. Further, Johnson advocates a pragmatic approach that rises above any positivist dogma of physical sciences and normative dogmas of social sciences. In this paper, we strive to combine the best of know-how in information systems and the agricultural domain.

The rest of the paper is as follows. Section 2, *Information and Agriculture*, covers the significance of information to the agricultural sector with an overview of agricultural information systems. Section 3, *Overview of Ecosystems* describes a variety of ecosystems. Section 4, *Analysis of Indian Agriculture Sector using Ecosystem Paradigm* analyses the status of the Indian Agricultural Sector along ecosystem dimensions. Section 5, *Analysis of Indian Agricultural Ecosystem using Tantra Framework* describes how Knowledge-based Tantra Framework can be used for analysis using descriptive, normative, and transformative viewpoints. Section 6, *Discussions* compares the proposed approach with prevailing alternatives. Section 7, *Conclusions* calls for defining a transformative vision and developing new mental maps for the Indian Agricultural Sector and concludes the paper.

## 2. Information and Agriculture

The importance of information to agriculture is well-recognized for the last many decades. The world bank discussion paper by William Zep [8] recognizes this as "information poverty" particularly affecting the developing countries. This included (i) difficulty of rural populace to get important information such as market produce prices and bulletins about pest infestations; (ii) literacy barriers; (iii) communication barriers; (iv) lack of systems to make productive use of indigenous knowledge and preserve it for future generations; (v) limited access for field workers to information about the population they are to serve and to current research findings (vi) lack of access to up-to-date information from the field to researchers (vii) inadequate information to government officials who have to make plans and decisions as well as monitor the progress of plans and consequence of decisions; (viii) inadequate provision of quality communication services to rural areas; (ix) absence of mechanisms to exchange information among the developing countries. The paper also envisaged a 'Rural Information Utility'. Some of these issues continue to be relevant.

Susan C Harris [9] identified issues in the distribution of agricultural information in developing countries such as geographic isolation, intellectual isolation, organizational isolation, and cultural isolation. Here intellectual isolation impedes multiple disciplines from working together. Social isolation makes people follow their social groups limiting the ability of Governments and experts to influence them.

Galtier et al [10] discuss the evolution of the Agricultural Market Information System (MIS) in developing countries and list the objectives as (i). regulation of the price of grains and avoid famines; (ii) dissemination of market information to the market players to encourage efficient and transparent trading. These MIS's catered to public information and operated at the national level. They faced challenges such as unreliable information, long dissemination timelines, no analysis, failure to assess the actual use made of the information provided, lack of institutional incentives for innovation to meet user needs, administrative rigidity, and short-lived, project-based funding. Over time the Second generation MIS made better use of communication technologies and focused on market transparency, targeted specific geographies, and covered wider produce as well as credit information. This MIS still fell short in informing on Public Policy as well as the ability to monitor and evaluate changes on the ground. Financial Sustainability is a critical concern for these MISes.

Dr. Sahadev Singh[11] in his paper describes chain of components that make up the Agricultural Information System which include (i) information organizations that generate and process information; (ii) information platform that enables dissemination and exchange, (iii) information bus/pathway that transports information between the information platform and user community; (iv) information and knowledge intermediaries that intermediate either by localizing or globalizing information; and (v) a user community that is not geographically defined but forms based on common needs, objectives, values, ccommodités, écorégions, disciplines, etc. The role of ICT is primarily to further enable, enhance and enlarge the "information spaces" and the user communities and enable learning within the user communities. They recommend strengthening national-level bodies as well as local extensions that reach out to farmers,

Julio A. Berdegué and Germán Escobar [12] recommend that we have to start from understanding the multi-dimensional and heterogeneous nature of rural poverty and then repurpose Agricultural innovation to address diverse groups with varied contexts and differing innovation affinity. They expect agricultural innovation to help in expanding rural non-farm economies.

When it comes to information delivery to farmers, agricultural extensions typically funded by Governments, play an important role. Adhiguru et al [13] did a study of the Agricultural Extensions in India and recommended a pluralistic approach to information delivery. Msoffe and Ngulube [14] emphasized the need to understand the target community before embarking on information dissemination. Table 2 below illustrates key concepts they use from DeLone and McLean [15]. The QUF is due to Mardis[16].

**Table 2 Designing Successful Information Systems**

| Model | Factors |
| --- | --- |
| **Information System Success Model** | Intention to use, Use, User satisfaction, Net benefit |
| **Information Quality** | accuracy, relevance, precision, reliability, completeness, usefulness, currency preferred format. |
| **Service Quality** | support users receive from the service provider |
| **System Quality** | accessibility, ease of use, ease of learning, intuitiveness, system reliability, system flexibility, sophistication, and response times. |
| **System Use** | amount, frequency, nature, extent, and purpose of the use. |
| **User Satisfaction** | how the user feels about the whole experience with the system. |
| **Net Benefits** | how much the information system adds to the success of the individual, group, organization, industry, or nation |
| **Quadratic Usage Framework (QUF)** | Dynamics of usage based on personal and environmental factors |

Neil Patel et al. [17] concluded that the farmers are likely to be influenced by their peers and online communities and prefer voice-based delivery mechanisms. Umunna Nnaemeka Opara [18] discussed the effectiveness of agricultural extension in Nigeria. Vidanapathirana [19] stress on the importance of agricultural information system and the latest research to farmers.

In India, special attention was paid to the development of an agricultural research infrastructure immediately after Independence. The Indian Council of Agricultural Research (ICAR) acts as a repository of information and provides consultancy on agriculture, horticulture, resource management, animal sciences, agricultural engineering, fisheries, agricultural extension, agricultural education, home science, and agricultural communication. It has the mandate to coordinate agricultural research and development programs and develop linkages at national and international levels with related organizations to enhance the quality of life of the farming community. ICAR has established various research centers to meet the agricultural research and education needs of the country. Then there are bodies such as AGRIS, AGRICOLA, and many others who address the same function at the international level.

Rao [20] has done a study on arriving at the right framework for implementing ICT in the agricultural sector in India. The proposed framework is visualized from the two perspectives (i) the rural incomes and livelihoods perspective at the farm level, and (ii) the sustainability perspective at the regional level. He recognizes the role played by the private sector here. ITC made use of a network of eChoupals or village-level information centers that enabled farmers to sell directly to the Company at prevailing market price, by bypassing the Mandi system but continuing to involve commission agents in logistics. The model also leveraged ICT to aggregate demands for farm inputs, allow free access to knowledge on crop management and expert advice, weather forecasts, and information on prevailing prices in local and global markets. There are other ICT platforms in India such as iKisan.com by Nagarjuna Group iVillages, n-Logue, and GyanDoot (operated by Madhya Pradesh Government).

The second aspect Rao [19] covered was sustainable development. For example, the development of large-scale surface irrigation initially raises crop yields, but in the long run, it can lead to waterlogging and build-up of soil salinity which can reduce land productivity. Intensive use of fertilizers and pesticides can lead to surface and groundwater pollution. Methane emissions from rice fields and livestock manure contribute to increased concentration of greenhouse gases leading to climate change and associated uncertainties and impacts on agricultural production. To address this Rao proposes setting up an institutional environment of a spatial data services network in the public sector that permits distributed access for integrating natural resources and socioeconomic data at local/regional/national/global scales to address sustainability concerns.

Another area farmers particularly need assistance in is how to manage uncertainties and risks. There are several studies related to this. Pažek and Rozman [21] examine decision-making under uncertainty in agriculture and propose alternatives. Nitsenko et al. [22] have done a study for the agricultural sector in Ukraine and propose a practical approach that is affordable and usable by agricultural entrepreneurs. Vermeulen et al. [23] explore addressing uncertainty in adaptation planning of agriculture.

In summary, the major thrust in agricultural information systems has been disseminating information and knowledge to farmers to help them improve productivity, access market price information and promote sustainable farming, etc. The public policy dimension and social/welfare dimension which predominate Indian Agriculture Sector are not adequately addressed. In addition, India has issues to tackle such as ecological degradation in states like Punjab and Haryana. To enable multi-dimensional analysis, we need to make use of the ecosystem paradigm.

## 3. Overview of Ecosystems

Peltoniemi, M. and Vuori, E. K. [24] give an overview of different ecosystems in their paper that compares business ecosystems with other ecosystems such as biological, industrial, the economy as an ecosystem, digital and social ecosystems. World Resources Institute [25] viewed the ecosystem as a solution to particular challenges in life. Kaufman [26,27] states that natural ecosystems are not totally connected but "typically, each species interacts with a subset of the total number of other species; hence the system has some extended web structure", resembling a huge scale-free graph. Lotka [28] and Volterra [29] represented the biological ecosystem using a predator-prey model and proposed a set of first-order non-linear differential equations to capture the dynamics of the biological ecosystem. Rothschild [30] draws an analogy between biological systems and economic systems by saying "Like the organisms and species that make up the global ecosystem, the world's firms and industries have spontaneously coevolved to form a vast living ecosystem.". Vuori [31] suggests that societies should strive to enhance knowledge intensity at the ecosystem level. This will require a degree of cooperation combined with competition.

The business ecosystem concept was introduced by Moore[32] in his article on predators and preys in Harvard Business Review. Peltoniemi, M. and Vuori, E.[24] built on these ideas and defined the Business ecosystem as **"**a dynamic structure which consists of an interconnected population of organizations. – Business ecosystem develops through self-organization, emergence, and coevolution, which help it to acquire adaptability. In a business ecosystem, there is both competition and cooperation present simultaneously". Thus, the business ecosystem is a concept that emphasizes interdependence, self-organization, emergence, and co-evolution of agents. Business ecosystems are compared to complex evolving systems, which can be simulated using agent-based modeling. In complex systems, there are emergent properties, which cannot be anticipated based on knowing the separate components of a system.

In a business ecosystem, companies "co-evolve" around an innovation, working cooperatively and competitively to support new products and satisfy customer needs. Here some companies play the lead role and are referred to as the keystone species. These companies have a strong influence over the co-evolutionary processes. Moore also recognized decentralized decision-making and self-organization within business ecosystems. In a business ecosystem, companies go through a lifecycle that consists of four stages: birth, expansion, leadership, and self-renewal/death. Iansiti and Levien [33] introduce four different roles that organizations can take in business ecosystems: keystones, niche players, dominators, and hub landlords. Dominators and hub landlords are the kinds of organizations that attract resources from the system but do not function reciprocally. Power and Jerjian[34] compare the flow of energy in natural ecosystems to the flow of capital in business ecosystems, both should be used efficiently for ecosystems to prosper.

Pagie [35] discusses three types of coevolution, namely competitive, mutualistic and exploitative. *Competitive coevolution* occurs between species which are limited by the same resources. In that case, the organisms are forced to change in such a way that they can either take advantage of that resource more efficiently or acquire the resource more efficiently. *Mutualistic coevolution*, on the other hand, comprises reciprocal relations where all the participants benefit from the interaction and change in the direction of better compatibility. *Exploitative coevolution* comprises relations where all the participants do not benefit from the interaction. A price war is an example of competitive coevolution, as is also the development of competing technologies. Mutualistic coevolution may be observed when organizations develop capabilities for cooperation and complementation to compete with a third party. Exploitative coevolution may be detected in a situation where an organization is significantly more powerful than the others. This could happen in the context of a large corporation and its suppliers.

In a business ecosystem context, self-organization implies the absence of a central or outside controller. Organizations make their own decisions based on imperfect, perhaps locally restricted knowledge that they possess. The formation of a business ecosystem is a process, where participants are gathered voluntarily and without an external or internal leader. Goals are set in local interactions, where companies negotiate and create a new order. The system has the freedom to organize according to the needs and capabilities of the organizations. Self-organization is enabled by the market economy system. In any real-life business ecosystem, however, there are interventions by the public sector, such as business subsidies, import duties, and publicly-funded development projects. These can be seen either as inhibiting self-organization or as creating enabling structures for self-organization.

Smith and Stacey [36] dwell on emergence. As emergence takes place, the behavior observed at the macro level is not obvious while examining the behavior at the micro-level. "Emergence means that the links between individual agent actions and the long-term systemic outcome are unpredictable". They also state that emergence makes it impossible for one actor to control the whole system, whether inside or outside the system. Emergence is a phenomenon that arises from organization-level motives and actions that lead to unpredictable and even surprising population-level behavior. It is induced by each organization's restricted knowledge of its environment, of its options, and the outcomes of those options. Thus, the phenomenon may be amplified in the population and result in an unanticipated situation. In Figure 1, coevolution, self-organization, and emergence are presented.

The fourth property of business ecosystems is adaptation. According to Holland ]37,38] adaptation generates "structures of progressively higher performance". There are three components associated with adaptation: the environment, the adaptive plan, and a measure of performance/fitness. Adaptation can be criticized for the passive role of the environment. Adapting always means adapting to something, and it incorporates the thought that the adapting unit is not capable of affecting its environment. When the environment changes, a business ecosystem adapts to changed conditions by emergence, co-evolution, and self-organization. Self-organization, emergence, coevolution, and adaptation are also key concepts of complexity theory. Peltoniemi and Vuori [39] elaborate on these phenomena in much greater detail.

As ecosystems expand and competition becomes harder, the businesses/organizations have to develop "dynamic capabilities". The dynamic capabilities are signature processes/offerings which give an edge to a business, in contrast to ordinary capabilities which competitors can easily emulate

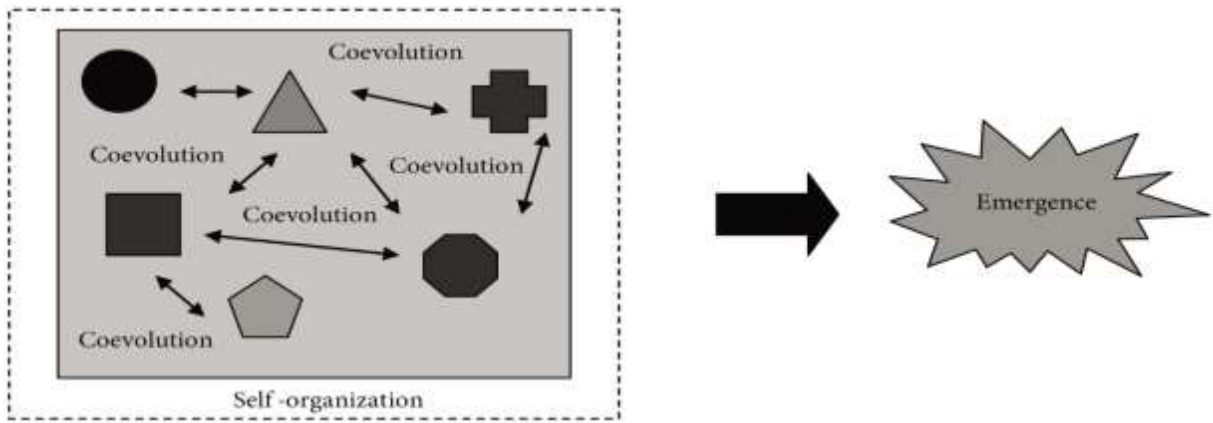

**Figure 1: Ecosystem Phenomena**

Finally, we need an ecosystem to manage ecosystem information. Marrow et. al. [40] define Information Ecosystem as "an interacting network of information producers and consumers, where interactions are typically facilitated by software agents or information agents when they are focused on manipulating information. In information ecosystems, the emphasis is on interactions between entities (information producers and consumers) in an environment of continuous change, caused by commercial, political, social and technological developments"

## 4. Analysis of Indian Agricultural Sector using Ecosystem Paradigm

Even though most activities have moved from the public to private sector in many countries, Governments' role has not diminished. Rather Governments now have a critically important role in managing ecosystems, which is a lot more complex. The agricultural ecosystem can be considered as an ecosystem of ecosystems comprising of ecological, social, economic, business, welfare, and industrial ecosystems.

An important dimension of the agricultural ecosystem is sustainability. Countries such as India for millennia practiced traditional farming using live-stock and natural ingredients and were global leaders in practice going back millennia. As the need to feed the millions and ensuring that those who are involved in farming have decent incomes, the use of fertilizer and pesticides as well as high-quality seeds became commonplace. Artificial Fertilizers were first created in the 1940s. Similar advances occurred in chemical pesticides in the 1940s. The Fertilizers are due to Justus Freiherr von Liebig, who has been described as the "father of the fertilizer industry" for his emphasis on nitrogen and trace minerals as essential plant nutrients, and his formulation of the law of the minimum, which described how plant growth relied on the scarcest nutrient resource, rather than the total amount of resources available.[41]. In India, the widespread use of fertilizers while beneficial in the short term, had serious longer-term side effects such as soil compaction, erosion, and declines in overall soil fertility, along with health concerns about toxic chemicals entering the food supply.

As inorganic farming based on fertilizers was becoming commonplace, another set of scientists were working on "organic farming". Albert Howard and his wife Gabrielle Howard, accomplished botanists, ventured to improve traditional farming methods in India. They combined methods from their scientific training; with traditional Indian farming methods, and developed protocols for the rotation of crops, erosion prevention techniques, and the systematic use of composts and manures. They called their technique Organic Farming, borrowing the term from Northbourne. Another early contributor to this movement was Ehrenfried Pfeiffer who termed it bio-dynamic farming. Organic Farming however tends to require expensive inputs and works out only when consumers are willing to pay a premium. Wikipedia [42] has covered the Indian connection to organic farming. Suryatapa Das et al have covered traction organic farming has received in India [43].

In recent years there is a reemergence of a variant of the traditional farming technique which is popular in India by the name Zero-Budget Natural Farming (ZBNF), pioneered by Subhash Palekar, which can be considered close to agroforestry/agroecology. This method makes use of only natural ingredients and requires no expensive farm inputs. The four pillars of ZBNF are (i) Jeevamtra (a fermented microbial culture that provides nutrients, and catalyzes activity of microorganisms/earthworms in the soil. This makes use of Cow-dung and urine); (ii) Beejamrutha is a treatment used for seeds, seedlings or any planting material again using similar ingredients; (iii). Acchadana – Mulching using Soil Much, Straw Mulch and Live Much, which avoids tilling (Mulch is simply a protective layer of a material that is spread on top of the soil) and (iv) Whapasa moisture, which is a condition where the soil has requisite air molecules and water molecules, thus obviating the need for irrigation. There are other champions of natural farming such as Masanobu Fukuoka from Japan. Details on Zero-budget Natural Farming are given here in this FAO[44] article. A case study on Zero-budget Natural Farming is available here [45}.

The second important dimension of the Agricultural Ecosystem is social. The structure of society does influence agriculture greatly. Many farmers engage in farming as a family venture. Others engage farmworkers. Some are tenant farmers who till other's land. There are also cases where the farmers form cooperatives. Then there are farm aggregators, intermediaries, and feudal landlords who have been engaging in their professions going back many generations. There are also socio-economic aspects such as education level which play a role in adopting newer technologies. Social structure also is key as some significant farm revolutions had a social basis. The white revolution was successful thanks to the community of Patels and the Green revolution due to the enthusiastic participation of Jat Sikhs.

The third important dimension of the Agricultural Ecosystem is economic. Nearly every Government provides support to Agriculture in some form or other. In India, this leads to huge outlays to support farming through the life cycle. Huge amounts of grains are procured yet times even in the absence of demand. Then there are all kinds of subsidies that distort markets.

The fourth dimension of the agricultural ecosystem is business. A farmer can sell his produce to a trader, government to support its welfare programs, industry, or use the farm to generate collateral revenue. . Contract Farming is slowly becoming popular. Farmers can also use his farm as a solar farm, windmill farm, or leisure farm to boost agritourism.  The bargaining power of farmers may become less when they sell through intermediaries or to industry unless they organize themselves. With the emergence of E-commerce, farmers can sell directly to consumers in particular fruits and vegetables.  This gives rise to Digital Business Ecosystems.

The fifth dimension of the agricultural ecosystem is welfare. The government of India had PDS initially and then they refined it to cover Above Poverty Line and Below Poverty Line Categories. This was a few years back followed up by the Food Security Act which covers nearly 2/3 of the population putting huge fiscal strain on the Government. There are MGNREGA (Mahatma Gandhi National Rural Employment Guarantee Act) schemes to provide employment opportunities during lean seasons in villages. MGNREGA has also pushed the minimum wage as well as reduced the labour availability to farmers. The government bears huge costs to support its welfare programs to benefit both consumers and farmers.

The sixth dimension of the agricultural ecosystem is industrial. Some organizations like ITC, PepsiCo entered into contract farming. In addition, Super Market chains also like to procure directly from farmers

Keeping track of information associated with all these manifestations has resulted in an information ecosystem that has information producers, intermediaries, and consumers. Farmers need to be informed of state of the art in agricultural research and know-how. Government and other stakeholders, in turn, need to know the ground reality of farm inputs, practices, outputs, and associated risks and uncertainties.  Figure 2 below illustrates the agricultural ecosystem. Here the right circles are concerned about production end and the left circles about consumption/trade aspects.

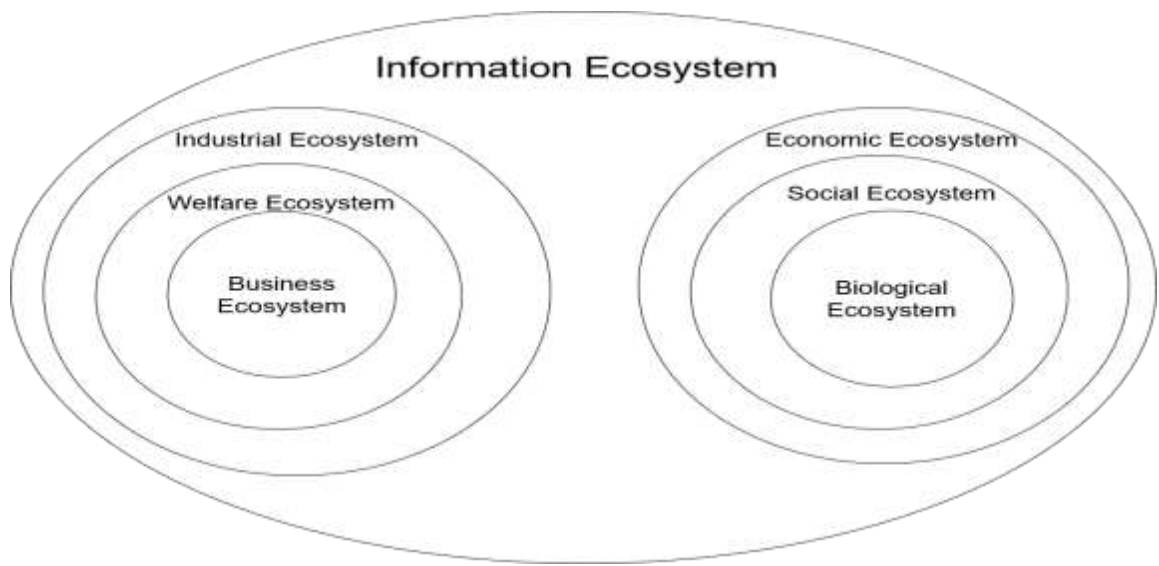

**Figure 2: Agricultural Ecosystem**

The issues in the Indian Agricultural Sector can be categorized along ecosystem dimensions as shown in Table 3 below.

**Table 3: Issues in Agricultural Ecosystem**

| **Biological/Natural Resources Ecosystem** | |
|---|---|
| **Water** | Water is not equitably distributed regionally. Water-deficit states like Punjab are growing crops such as Rice. Sugarcane also takes a lot of water. Dams have failed to solve the problem due to last-mile issues. Cheap power has meant over-drawing of water and depletion of water tables. Irrigation has helped certain states enormously while many others depend only on Monsoon |
| **Land** | Land is a scarce resource needed for industrialization as well as agriculture. While ideally, farmers should be free to sell their land, indiscriminate diversion of land from agriculture cannot be allowed. |
| **Soil** | The use of fertilizers in a non-judicious manner leads to erosion of soil. Organic Manures and Biocides also can cause soil erosion |
| **Health** | Pesticides increase the cost as well as spoil health. |
| **Seeds** | High-quality seeds need to be available at affordable prices to farmers |
| **Weather** | Periodic crop losses due to ill-timed rains, floods, and drought put farmers in distress |
| **Social Ecosystem** | |
| **Land Distribution** | Small and fragmented land holdings make farming unviable and reduce the bargaining power of farmers |
| **Economic Ecosystem** | |
| **Productivity** | Agriculture engages nearly half the population but accounts for less than 20 percent of GDP. Lack of Mechanization in farming methods puts Indian farmers at disadvantage. |
| **Debt** | Farmers take a loan from individuals as well as institutions and many cases unable or unwilling to pay back. |
| **Price-Support** | Many Government schemes such as Minimum Support Price helps rich farmer lot more than their poorer counterparts. Typically, rich farmers purchase from smaller farmers and sell to the Government. |
| **Return on Investment** | The profitability in crop cultivation from public irrigation hardly matches the opportunity cost of public irrigation. Public irrigation projects are very capital intensive but the financial outcomes hardly make justice to investment. |
| **Business Ecosystem** | |
| **Supply-chain and Agricultural Marketing** | Farmers are restricted to sell in their local Mandis leading to an inefficient supply chain. Studies have shown that farm produce sometimes changes ten times without making any value-addition. Both producers and consumers feel short-changed. |
| **Market Separations** | Farmers in many cases are forced to sell in the local Mandis. This itself is a barrier. Due to a variety of reasons, farmers may be forced to sell to aggregators |
| **Trading** | As a part of the procurement function, there is a need for people/institutions who do quality assessments of the crops. Traders would also like to avail of insurance facilities when they are trading with unfamiliar parties/locations. |
| **Lack of Quality Control** | In many cases, there is inadequate quality control of grains as it moves through the supply chain. At times inferior grains are purchased and at other times their quality declines as they move through the supply chain. Quality is a key issue to be addressed to liberalize the sector |
| **Scarcity of Capital** | This forces farmers to use credit leading to a debt trap |
| **Shortage of Transport facilities** | This makes farmers dependent on intermediaries and at times forces them to sell at loss. |
| **Welfare Ecosystem** | |
| **Market Distortions and Supply Shocks** | As Governments buy huge quantities of rice and wheat it affects cropping patterns. A year of drought may be followed by a bumper harvest. |
| **Storage and Distribution** | To support farmers, Governments, procure lot more grains than they need. This needs huge storage. Lack of adequate storage means grains are stored in the open and lead to the rotting of grains. Currently, only 10 percent of the requirement is met as far as storing wheat in steel silos is concerned |
| **Information Ecosystem** | |
| **Lack of quality information** | It is much easier to target schemes to farm owners than tenant farmers or farm workers due to a lack of information. As a result, those who need a benefit may not get it. |

An alternate view of the ecosystem with participants, stocking points, and flows is shown in Figure 3 below. Table 4 covers the participants in the agricultural ecosystem.

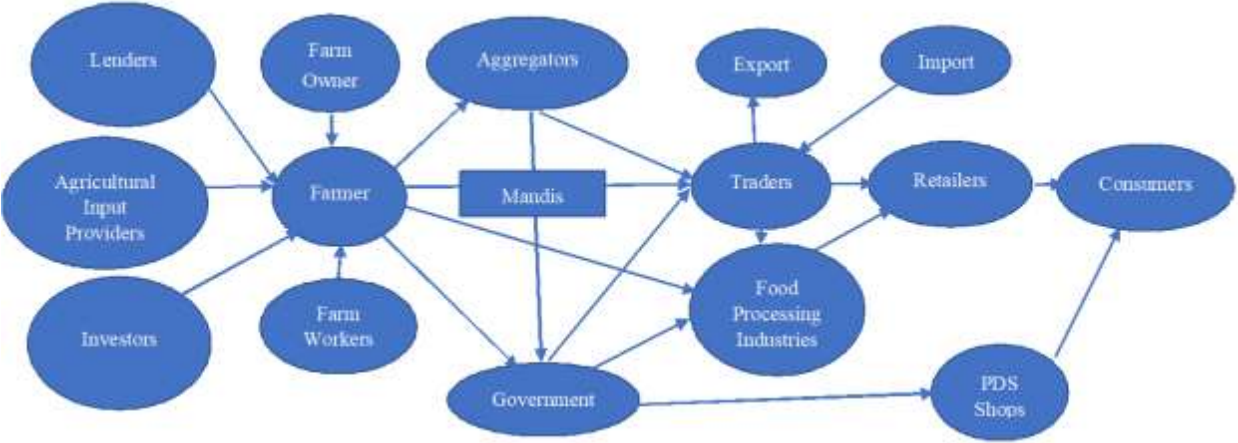

**Figure 3: Flows in Indian Agricultural Ecosystem**

**Table 4: Participants in Indian Agricultural Ecosystem**

| | |
|---|---|
| **Biological/Natural Resources Ecosystem** | |
| **Agricultural Input Providers** | Farming requires good quality seeds, soil, and water. Modern farming relies heavily on fertilizers and pesticides. Farmers in turn depend on power at low rates for watering their fields. |
| **Research Institutions & Extensions** | The government has set up specialized research institutions as well as agricultural universities to guide better farming methods. |
| **Social Ecosystem** | |
| **Farm Owners** | Farm owners may choose to do farming on their own or give it to a tenant Farmer to do the farming. As a part of land reform, large amounts of land were transferred from farm owners to tenant farmers by Government, disregarding property rights. About 85% of farmers now have smallholdings. |
| **Farmers** | Farmers may be Farm Owners or Tenant Farmers. Farm owners may involve their own family in farming and/or engage Farmworkers. |
| **Farmworkers** | Farmworkers work on farms. They do not necessarily get benefit if the price of produce increases |
| **Farmer Co-operatives** | In the case of certain crops such as sugar, farmers have established co-operatives. The idea is to give a fair price to farmers. These cooperatives are very popular in states like Maharashtra, Punjab, Karnataka, etc. |
| **Economic Ecosystem** | |
| **Government** | Government plays an all-pervasive role in the agricultural ecosystem. For example, several arms of the Government decides on Support Prices of 20-30 odd products. Agriculture costs and Prices commission under Ministry of Agriculture recommends MSP for 24 crops. |
| **Regulators** | APMC act regulates trading and farmers have to sell at designated Mandis. E-NAM which is newly introduced will also need to be regulated. |
| **Business Ecosystem** | |
| **Consumers** | These Consumers purchase from the open market and are affected by price volatility |
| **Lenders** | These are banks, micro-finance companies, and local money lenders. Local money lenders occupy a significant portion. There are specialized banks such as NABARD. |
| **Investors** | A farmer may invest his own money or take it from other investors. Contract Farming companies and Food Processing companies invest in agriculture and then buy the produce from farmers. |
| **Mandis** | Farmers typically sell their produce in Government approved Mandis (Markets) to traders at those Mandis. In the case of certain produce, farmers are forced to sell at Mandis by law. |
| **Intermediaries (Arthiyas)** | Intermediaries aggregate from Farmers and sell to Government. Intermediaries also offer credit to farmers at times leading to monopsony. They are commission-agents/middlemen/brokers. |

| **Traders** | Traders buy from Farmers at Mandis or buy directly from farmers and then sell at Mandis by aggregating produce. Only authorized persons can sell or buy at Mandis. |
|---|---|
| **Retailers** | The retailers buy from wholesale traders and fulfill requirements in the neighborhood. Retailers may also be large stores that are in multi-brand retail. |
| **Transporters** | They are an integral part of the ecosystem. If the transport cost is too high then farmers are forced to sell to middlemen or at discount at Mandis. |
| **Importers** | Certain agricultural produce is imported in large quantities when local production cannot meet the demand. For example, palm oil is generally imported from South-East Asia. |
| **Exporters** | Exporters focus on exporting items in demand in export markets. At times exporters are restricted from exporting when local production falls below and prices increase. For example, Basmati Rice to vegetables and fruits. |
| **Welfare Ecosystem** | |
| **Beneficiaries** | Consumers who rely on the public distribution system. |
| **Industrial Ecosystem** | |
| **Food Processors** | Companies such as Pepsico, Britannia, and ITC process food and then distribute it to their retail chain and super-markets |
| **Supermarket chains** | Corporate entities such as Reliance in India run their retail outlets. |

To transform the Indian Agricultural Sector, it is critical to reducing the mutual dependence between the sector and the Government. In this section, we explore an approach to achieve that. The government of India invests in agriculture in a variety of ways. The government has invested in agricultural research. The government regulates the agricultural sector heavily. Government disallows farmers from selling their goods except through authorized markets. Government disallows hoarding and cartelization. Government has export restrictions when prices go up. Undoubtedly there is a need to do a retrospective on the footprint of Government which has steadily expanded over the last seven decades. The role currently played by Government is depicted in Figure 4 below. Table 5 provides details on the role of the Government in the Agricultural Sector.

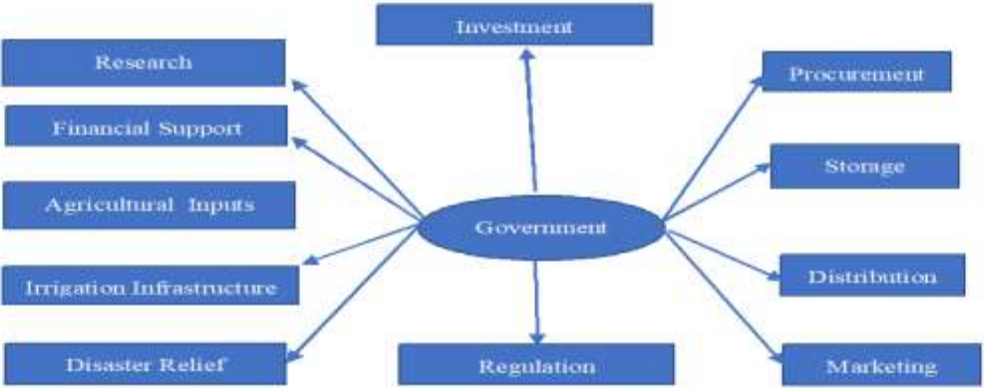

**Figure 4 Government's role in Indian Agricultural Sector**

**Table 5: Government's Role in Indian Agricultural Sector**

| | |
|---|---|
| **Biological/Natural Resources Ecosystem** | |
| **Irrigation Facilities** | Government invests in irrigation facilities and last-mile connectivity of irrigated water to farmers. |
| **Agricultural Research** | The government has set up many organizations and universities to promote agricultural research |
| **Agricultural Extensions** | Dedicated departments and personnel interact with farmers and advise them on best practices in farming |
| **Economic Ecosystem** | |
| **Input Support Means support** | Subsidized Fertilizers in particular Urea provided to farmers. The subsidy money however goes to the Fertilizer companies. This may result in many inefficient producers continue to survive without adopting the latest technologies. |
| | Subsidized/Free Electricity. Here State or Central Governments bear the cost of power |
| | Seeds at reduced prices to Farmers |
| | The government also supports the installation of solar pumps and the like in farms. |
| | Telangana Government has a Raitabandhu scheme where Government pays cash to farmers depending on acres he tills/owns, which they can use to buy seeds, fertilizers, and pesticides. |
| **Income tax relief** | Any income earned from proceeds of agricultural activities is exempt from income tax. |
| **Procurement** | The government has set up the Food Corporation of India to centrally procure food grains to enable public distribution of food grains and to build buffer stocks. |
| **Policy-based interventions in Agricultural Market** | Government disallows the free flow of agricultural -goods. Both export and import are regulated. Government sets Minimum Support Prices which are at times more than market prices. The government's support may mean farmers do not focus enough on the productivity of farming. This makes the sector less competitive. In addition, MSPs direct crop choices thus causing deficits of items like pulses, fruits, and vegetables and excess of cereals or any item that Government supports. Farmers have less incentive to move to organic and sustainable agricultural practices. Government disallows hoarding and cartelization and limits by making use of the essential commodities act which sets stock limits. These issues are to be addressed by the new farm laws. |
| **Business Ecosystem** | |

| **Price Support** | The price support is in the form of Minimum Support Price mainly to rice and wheat. There are 23 crops Government tracks and more and more crops are included in the minimum support price scheme. These Prices may be higher than the market at times. Madhya Pradesh Government has come up with an alternative scheme called Price Deficiency where Government instead of procuring grains pays the difference between Market Price and Average Sale Price (ASP). |
|---|---|
| **Loan Support** | Crop loans are given at a lower rate where Government picks up part of the interest using interest subvention schemes. In many cases, loans are waived. UPA Government waived Rs. 71,000 Crores in 2009. |
| **Risk Management** | The government pays most of the premium for crop insurance. The government pays when crops are damaged or lost, directly or through insurance schemes. |
| **Storage** | Government invests in storage facilities in the form of Steel Silos to store procured grains, thus enabling the Government to procure in advance. |
| **Marketing** | The government has set up markets for agricultural produce which go by the name APMC |
| **Financing** | The government has set up NABARD (National Bank for Agricultural Development and Rural Development). |
| **Welfare Ecosystem** | |
| **Income Support** | Government should strive to ensure market prices to be high enough to ensure desired income to farmers through appropriate price support. There are two schemes for income support: (i) A2+FL +x% factors all input costs farmer directly bears and cost of family labor (ii) C2+x% factors all costs included in A2+FL as well as imputed costs for land and interest on capital. The government generally uses the first scheme. |
| **Distress/Disaster Support** | The government will intervene as appropriate in case of floods, droughts, and other emergencies. |
| **Livelihood Support** | Recently Government of India announced a Rs 6,000 annual payout to poor farmers. This is going to cost Government 75,000 Crores benefitting 14.5 Crore farmers. |
| **MGNREGA** | This provides work opportunities for up to 100 days in Rural Sector. Even though anybody can make use of it. This is particularly helpful to farm workers during the lean season. |
| **Public Distribution System and Food Security Act** | Government buys Food Grains at MSP and then provides them to people at very low prices. In the case of the BPL (Below Poverty Line) category, it is almost free. In the case of the APL (Above Poverty Line) category, they have to pay a nominal rate. Up to 30 Kgs of food grains are made available to a family. Under the Food Security Act, the coverage is increased to 66 percent of the population. |

| Industrial Ecosystem | |
|---|---|
| **Regulation** | Government is responsible for providing the legal framework for contract farming and any other involvement by industry in the agricultural sector. Regulation should not be heavy-handed. |
| **Dispute resolution** | Government has to specify the mechanism for dispute resolution which is speedy and not cumbersome. |

We can next represent the reforms underway or being envisaged in the Indian Agricultural Sector along ecosystem dimensions. See Table 6 below.

### Table 6 Reforms in Agricultural Ecosystem

| Biological/Natural Resource Ecosystem | |
|---|---|
| **Organic Farming** | Organic farming fetches higher prices and it is environment friendly. Up-front investment in organic farming is far less and depending on the yield it may even fetch better returns. Sikkim has declared itself to be the first 100% organic state in India and won global recognition. There is a lot of export potential for organic produce. |
| **Zero-Budget Natural Farming** | This is a revival of traditional Indian farming technique which relies upon cow dung and vegetable residue   ZBNF is environment friendly even compared to organic farming. It does not involve buying organic compost from any of the companies and even avoids tilling. |
| **Reforming Agricultural Practices** | Farmers can be incentivized to shift to better farming practices through knowledge-sharing. Precision agriculture makes use of optimal quantities of fertilizer. |
| **Use of Science and Technology** | The use of Biofortified food can move India from food security to nutrition security. The government also needs to carefully evaluate the introduction of Genetically Modified Foods. Use of Solar Energy, Wind Energy in farming as well as adopting drip irrigation methods. |
| **Social Ecosystem** | |
| **Cooperative Sector** | The dairy revolution happened due to cooperatives.  Cooperatives can be promoted to a greater degree in agriculture. ZBNF is currently led by farmer groups |
| **Economic Ecosystem** | |
| **Reducing Centralized procurement** | The State Governments can procure food grain directly from each other.   FCI can just maintain adequate reserves to tackle food emergencies. So far, this proposal has not got traction. |
| **Supporting Rural Economy** | The agricultural sector cannot sustain itself in villages if villagers steadily leave for cities and towns. Local economies should be supported and blind pursuit of urbanization should be stopped |
| **Doing away with an income tax exemption for Agricultural income** | The tax exemption helps only rich farmers. On top of that many use it to declare non-agricultural income as agricultural income and evade tax. This provision disempowers the agricultural sector to be self-sustaining. It drains resources from the economy at large to support the agricultural sector. As of now, this proposal finds few takers from the political class. |
| **Business Ecosystem** | |

| **Electronic Nationwide Agricultural Market (ENAM)** | This enables nationwide trading of agricultural goods, by inter-connecting Mandis across the nation. Traders pay fees for a single license to trade nationally. The government needs to assure trust, quality as well as safe and secure transit of goods. |
|---|---|
| **Welfare Ecosystem** | |
| **Imaginative solutions to financial distress** | Politicians need to move from price support policies or loan waivers to income/investment support on a per-acre basis. The payment can happen using Direct-Benefit-transfer schemes. This can be both good politics and economics |
| **Reform PDS and Food Security regime** | The government should move to Flexible Social Benefit Plans [11] and optimize the procurement of food grains. As districts become more prosperous, people should be given greater choices. |
| **Industrial Ecosystem** | |
| **Contract Farming** | Contract Farming has taken off in certain states. Yet the contribution from Contract Farming is no more than 2%. |

If we were to take stock of the progress India made in agriculture, we can see that there was a lot of focus on Economic and Welfare ecosystems. The involvement of the Government and fiscal burden on the Government should reduce. There are opportunities to promote organic and natural farming in the biological ecosystem. The cooperatives sector is a big opportunity to be tapped in Social Ecosystem. In general, empowering farmers is the need of the hour. In the Business Ecosystem, trading restrictions should be eased. Greater involvement of Industry via contract farming and direct purchase by super market along with investment and facilitation by the private sector is desirable.

In the biological/natural ecosystem, it is extremely important to reduce the usage of fertilizers. Currently, subsidies are directly paid to the fertilizer companies. If the same amounts are given to farmers as benefits, they may choose to procure only what they need and be more open to other alternatives. This can also incentivize fertilizer producers to adopt more efficient and environmental-friendly methods of production.

In the Economic Ecosystem, we need to study how farmers respond to risks and uncertainties. In practical terms risks can be associated with actions where farmers have a choice such as a crop mix. Farmers may choose an alternative that has greater certainty even when the expectation value of a different alternative is higher. This in a way explains their preference to grow MSP crops even when other options may be more rewarding.

With regards to the welfare ecosystem. it is important to move beneficiaries of the public distribution system towards direct benefits and flexible benefits [46]. This can significantly reduce the costs associated with procurement, storage, and distribution. Making the social benefits flexible can help beneficiaries choose the benefits they need in quantities they need.

The current Business Ecosystem, where the farmers are required to sell at local APMCs, leads to Monopsony as well as a lack of space arbitrage, The small farmers may sell their produce to commission agents on whom they depend for credit leading to yet another Monopsony. In this setup, farmers had a lack of storage facility, inability to get a reasonable price, delay in sale and payment, high commission charges, and transportation problems, lack of transparency in quality-wise pricing of products. Model APMC act which made it easier for traders to participate remained on the paper. Even under the Model APMC act sales outside Mandis attracted Mandi fees and there was no integration of Mandis even within a state.

This was attempted to be addressed in 2016 by introducing E-NAM (Electronic National Agricultural Market) which linked APMCs across the state and at the national level where trading was allowed with a single license. This led to competitive bidding and better price recovery for producers. The platform also provided modern facilities for grading, price dissemination and transparency weighing, warehousing, electronic auction (e-bidding), and borrowing (loans in addition to crop loan) based on warehouse receipts. The effectiveness of E-NAM was studied by Arcot Purna Prasad et al. [47]. The aspects appreciated by the farmers under E-NAM were electronic weighing and grading, immediate credit of payment to the account, timely sale of products with good no. of bidders, transport facility to Mandi places, Mandis at nearby places, etc.

Another alternative for farmers is to sell their produce to Government which procures food grains to maintain buffer stocks at Minimum Support Price. Government announces MSP for 23 types of produce. These prices are generally higher than those prevailing at APMCs. However, if the farmers lack storage facilities their ability to profit from MSP is compromised and hence, they cannot benefit from time arbitrage. MSPs seldom are introduced in advance to enable farmers to make the right crop choices. Many farmers who cannot wait for procurement by Government end up selling to the middlemen or Arthiyas. Arthiyas also were providing credit to them leading to a monopsonist situation at micro-levels. Most of the MSP-based procurement happens only from Punjab, Haryana, and in recent years from Madhya Pradesh. From the viewpoint of the Economic Ecosystem, MSP is a very costly exercise to the Government. Many times, grains far above requirement are procured centrally via the Food Corporation of India. Then lack of adequate storage leads to spoilage of grains and diversion to make alcohol and animal feed. The issues related to MSP are discussed by Deodhar [48] as a backdrop to much-needed farm reforms.

The third alternative explored by many Governments is contract farming. In Punjab, there were apprehensions of ecological crisis due to water-intensive rice cultivation. Opportunities arose with Pepsico offering to procure Tomatoes and Chilies under contract farming to fulfill the pre-condition of entry into the Indian market. Then contract farming expanded to more states such as Gujarat and more food processors such as Mccain. The overall contract-farming-based procurement is less than 2% [48]. Vijay Paul Sharma [49] did a study on small-holder participation in contract farming. As per the study, contract farming benefited large and medium farmers more. Small farmers had greater difficulty in satisfying quality criteria. They were still dependent on existing channels such as middlemen and public institutions for credit.

Based on recommendations by the Agricultural Experts Government of India, passed 3 farm laws to reform the agricultural sector in the year 2020. Shankar [50] elaborated on the farm laws and concerns expressed by farmers. The laws are illustrated in Figure 5 below.

With the first farm bill, a farmer is free to trade with any merchant having a PAN number, anywhere in India. The second bill will allow a farmer to tie up with a large buyer, exporter & retailer and will help the farmer to assure price before their sowing. It will transfer market risks from the farmers to the promoter and give farmers access to higher quality seed, fertilizers, and pesticides as well as technical advice and risk-sharing. The third bill was essentially a product of the shortages era and it deregulates trade of major food=stuffs. The other important outcome expected is crop diversification beyond the water-intensive crops towards coarse grains and other foodstuffs.

**Three Farm Bills**

- The Farmer's Produce Trade and Commerce Bill, 2020 that allow farmers to sell theri harvest outside the notified Agricultural Produce Market Committee (APMC) mandis without paying any state taxes or fees.
- The Farmers Agreement on Price Assurance and Farm Services Bill, 2020, facilitates contract farming.
- The Essential Commodities Bill, 2020, deregulates the production, storage, movement and sale of several major foodstuffs, cereals, pulses, edible oils and onions, except in the case of extraordinary circumstances.

**Figure 5: Illustrating the Three Farm Acts 2020 Implemented by Centre Government [50]**

Some of the apprehensions expressed by the farmers regarding farm bills were (i) Contract Farming and freeing up markets will favour big private companies; (ii) Amendments to essential commodities act will allow black marketing (iii) Agri-business firms, wholesalers, processors, exporters and large retailers for farm services will manipulate market situations to gain at the cost of the farmers (iv) Laws on contract farming will put the land ownership of farmers at risk (v) Framers cannot protect their interests when exposed to large traders.

Satish Deodhar [48] did a study of Indian Farm Market Reforms attempted to be realized using the 2020 farm laws. His analysis makes a case for going ahead with the farm laws. Some of the collateral benefits he expects are stock-holding capability of the produce and more efficient supply chain and increased bargaining power of farmers in APMCs. Deodhar concludes that " Whenever there are welfare-inducing policy changes, gains to winners are much more than the loss to losers. However, if losers are few in numbers and much better organized, then they can lobby harder. Perhaps those who benefitted from the *status quo* are at the forefront of the protests. One more reason could be that quite a few state governments are likely to lose revenue since they will not get a market fee from transactions done outside of the APMC system. Add to that political considerations and foreign debilitating interests, and the farm market reforms have become a vexed issue".

Shoumitro Chatterjee and Mekhala Krishnamurthy [51] analyze the farm laws and argue that the laws are misdirected. They recommend increasing the number of Mandis in place of liberalizing the trading. One important point they make however is that the key issue is that farmers lack access to credit, inputs, storage, transport, and timely payments. Ajay Chhibber [52] covers farm protests and claims that the real solution is a second green revolution and to generate more employment in the non-farm sector. Many others oppose farm laws by spreading fear about the corporate takeover of agriculture or launch attacks on the present regime. Bhattacharya and Patel [53] lament about the apprehensions and anxieties in the minds of farmers that have led to the protests. Sudipto Mundle [54] argues that the farmers, on their part, can also do much to build FPO (Farmer Producer Organizations with countervailing power against the large contract-farming companies they fear. It is timely that the Government of India has chosen to form a new Ministry of Corporation which earlier was made part of other Ministries. Shikha Jha [55] in her 2007 article study concludes that decentralizing procurement from the Central to the State Governments can substantially reduce government costs with the little overall impact on producers, consumers, or trade.

There are also opportunities exploited by other countries in Leisure Farming and Agri-tourism which India can tap. Chang et al. {56} studied Leisure Farming in Taiwan. Further farm land can be used to generate Solar Energy and accommodate windmills.

Considering the complexities, criticality, and controversies the agricultural sector entails, having an all-encompassing information ecosystem that establishes truth and encourages transformation is the need of the hour.

## 5. Analysis of Indian Agricultural Ecosystem using Tantra Framework

Agricultural Ecosystem can be considered as Ecosystem of Ecosystems and we need an overarching Information Ecosystem to manage that. Some unique characteristics of this ecosystem are (i) phenomenal diversity of crops and multidisciplinary knowledge(ii) knowledge that spans the most ancient to the most modern (iii) complex supply chain and value chains (iii) need to manage risk, uncertainty and their aftermath and (vii) huge ecological, economic and societal impact. To meet the objective, we propose the use of the Ontological Knowledge-based Tantra Framework. The framework should be able to represent personal, social, institutional, and ecosystem information and contextualize it in geo-spatial, temporal, and governance contexts.

### 5.1 Ontological Knowledge-Based Tantra Framework

The objective of the Tantra Framework is to represent social information in a unified manner covering all aspects, linkages between aspects while accommodating different levels of detail using a process called reification. Tantra Framework extends Zachman Framework [57,58] by adding three additional columns namely relators, relationships, and separations to the prevailing six interrogative columns. Here the concept of relator and relationship are taken from Unified Foundational Ontology [59]. The framework thus is well-equipped to manage knowledge of any domain. Tantra Framework is implemented using Neo4J Graph which enables the visual representation of information as Knowledge Graphs.

The aspects of the Tantra Framework useful to express social information comprehensively are given below.

1. People/Households/Communities/Categories (Who)
2. Places/Addresses/Locations/Zones/Geo-spatial Constructs (Where)
3. Assets/Attributes/Entitlements (What)
4. Events/Seasons/Durations/Timelines/Temporal Constructs (When)
5. Processes, Interventions, Phenomenon (How)
6. Objectives, Measures and Metrics (Why)
7. Relationships between Aspects, Transactions, Affiliations, Social Networks, Ecosystems, Supply Chains, Value Chains, Social Circles (Relationships)
8. Relators/Intermediaries/Institutions (enable relationships and entitlements)
9. Separations (difficulty of establishing a relationship)

Table 7 below, illustrates the Tantra Framework. Each aspect in Tantra Framework is reified using perspectives at contextual(named), conceptual(defined). Logical(designed), physical(configured) and instantiated levels. See Tables 8 and 9.

**Table 7: Tantra Framework**

| Perspectives | Aspects | | | | | | | | |
|---|---|---|---|---|---|---|---|---|---|
| | Who | Where | What | When | How | Why | Relationships | Relators | Separations |
| Contextual (Named & Scoped) | Name | Name | Name | Name | Name | Name | Name | Name | Name |
| Conceptual (Defined) | Concept | Concept | Concept | Concept | Concept | Concept | Concept | Concept | Concept |
| Logically Designed | Relation | Relation | Relation | Relation | Relation | Relation | Relation | Relation | Relation |
| Physically Configured (Schema) | Network schema | Network schema | Network schema | Network schema | Network schema | Network schema | Network schema | Network schema | Network schema |
| Detailed/Instantiated | Unique ID | Unique ID | Unique ID | Unique ID | Unique ID | Unique ID | Unique ID | Unique ID | Unique ID |

Here any quantitative information is covered under the "Why" aspect as a Measure. For example, the kind of benefit, type of asset is covered under the "What" aspect but the value of the benefit or asset is covered under Why/Measure.

**Table 8: Reification of People Domain**

| Perspective | All people | Citizens | Residents | Aadhaar-Card Holders | PAN Card Holders |
|---|---|---|---|---|---|
| Named (Identified & Contextualized) | All the people within the identified context | People who are citizens within the identified context | People who are residents within identified context | People who have an Aadhaar Card within the Identified context | People Who have a PAN Card within the identified context |
| Defined (Conceptually Structured) | What makes one a member of this domain/role | What makes one a member of this domain/role | What makes one a member of this domain/role | What entitles one to an Aadhaar Card and where it is Mandated | What entitles one to an Aadhaar Card and where it is Mandated |
| Logically Designed | Related attributes that map to other aspects. | Related attributes that map to other aspects | Related attributes that map to other aspects. | Related attributes that map to other aspects. | Related attributes that map to other aspects. |
| Configured | Representation in Graph database as nodes and edges. | Representation in Graph database as nodes and edges. | Representation in Graph database as nodes and edges. | Representation in Graph database as nodes and edges. | Representation in Graph database as nodes and edges. |
| Instantiated | Instantiate with a unique ID. | Instantiate with a unique ID. | Instantiate with a unique ID. | Instantiate with a unique ID. | Instantiate with a unique ID. |

Table 9: Reification of other aspects

| Aspect | Examples |
|---|---|
| Address | Residential address, General Address/Location, Commercial Address, Institutional address, Address for civic amenity |
| Event | The birth event, Education Enrolment, Employment, Marriage, Retirement, Death, etc. |
| Asset | House Owned, Vehicle Owned, Land Owned, Business Owned, Stocks, etc. |
| Processes | Aadhaar Enrolment, Voter Enrolment, PDS Enrolment, Birth Registration, Property Registration/Lease |
| Artifacts | Aadhar Id, Voter ID, Public Distribution Service (PDS) card, Permanent Account Number (PAN) Card, Birth Certificate, Death Certificate, Property Sale/Purchase, Property Transfer/Ownership. Lease |
| Relator | Enrolment Agencies and Service Providers. |

Each aspect is identified at a high level with a name and context then reified at the conceptual level. The concepts then are unbundled logically. The logical representation of the aspect then is physically configured with any other properties, attributes. Finally, each aspect is instantiated by giving it a unique Id. In the case of personal identity, this may be a unique biometric Identity such as Aadhaar. Each aspect maps to domain and role. Thus, a person has Aadhaar Id as a natural person. As an Income Tax payer, he gets associated with PAN Id, and as a licensed driver, he may have a Driving License. Similarly, the properties may have unique property IDs which hold their uniqueness and validity in a given geo-spatial-temporal context. We can further categorize people based on their socio-economic characteristics

The strength of the Tantra Framework is its generic approach. An entropy construct was defined as an instrument of validation for the Tantra Framework [60]. Tantra Framework has been applied to a variety of scenarios related to E-Governance. Tantra Framework was used as Information Management Framework for Good Governance [61 and Knowledge Management Framework [62] and Network-based Framework for Managing Social Information [63]. Further Tantra Framework was used for Identity Management [64], Revenue Capture [65, Intelligent Social Banking [66], Balanced Development [67], and Balanced Growth [68]. Tantra Framework was positioned as a Transformative Framework for India's Electoral Democracy[69} and Social Sector[70]. In the Balanced Growth paper, we proposed a five-sector approach for the Indian Economy consisting of Knowledge, Manufacturing, Trade, Agriculture, and Services, which can be used to unbundle the Agricultural Sector and position it as a growth-oriented sector. In the Transforming Social Sector paper, we proposed the use of a Flexible Social Benefit Plan in place of the current Subsidy regime. This is particularly useful to reduce the Government's involvement in Food Grain procurement to what people truly need.

Next, we elaborate on how Tantra Framework can be used to model the agricultural sector using a descriptive viewpoint. Table 10 below describes the same. The Knowledge Graphs related to Tantra Agricultural Framework are depicted in Appendix A.

**Table 10: Modelling Agricultural Sector using Tantra Framework**

| Aspect | Participants/Entities Modelled |
|---|---|
| People (Who) | Farm-owners, Farmers, Tenant Farmers, Farm-workers, Small-Holdings Farmers, Medium Farmers, Large Farmers Rich Farmers, Aggregators, Traders, Retailers, Consumers, PDS Beneficiaries, APL/BPL beneficiaries., Money Lenders, Households, Consumers, Buyers, Sellers, MSP Beneficiary, APMC Farmer, Contract Farmer, Commission Agent, APMC Trader (Refer to Knowledge Graph (i) in Appendix A). |
| Plots (Where) | Geo-tags, Address and Locations, Geo-spatial attributes and classifications of Farm Plots, Zonal information of Localities, Villages, Districts, States, and Regions, |
| Events (When) | This may be announcements on schemes, payment of benefits, procurement, or any other event such as transfer of ownership of land. Life cycle events of farmers such as death and migration can also be linked and tracked. |
| Assets (What) | Own house, Vehicle, Tractor, Solar Pump, Land Holding, Dairy, Business Owned, Equity, Gold, Farm, Conventional Farm, Organic Farm, Leisure Farm, Solar Farm, Wind Farm (Refer to Knowledge Graph (ii) for types and kinds of farms) |
| Crops (What) | Rice, Wheat, Sugarcane, Pulses, Vegetables, Cereals, Coarse Cereals, Pulses, Oilseeds, Commercial Crops, Crops under MSP (Refer to Knowledge Graph no. (iii) in Appendix A) |
| Benefits received from the government (What) | Input Support, Price Support, Income Support, Insurance Support, Interest Subvention, Support, Compensation, Loan Waived (In case of farmers), Fertilizer Subsidy, Electricity Subsidy, Water Subsidy |
| | PDS, MGNREGA, Subsidies related to health, education, electricity, water, LPG, and many others |
| Incomes (What) | Income through Farming, Income through Farm Lease, Farm Labour Income, Cottage Industry, Animal Husbandry, Non-Farm Income, Money Lending |
| Credit/Debt (What) | Credit/Debt from Commission Agent, Trader, Banks, Money Lenders, Other forms of credit. |
| Knowhow (What) | Traditional Farming Knowledge, Natural Farming Knowledge, Innovation Affinity, Water Conservation Knowhow |
| Processes (How) | Input Support, Price Support, Income Support, Insurance Support, Interest Subvention, Support, Compensation, Loan Waived (In case of farmers), Fertilizer Subsidy, Electricity Subsidy, Water Subsidy. |
| Farming Process (How) | Modern Farming, Organic Farming, Natural Farming, Contract Farming, APMC, MSP, Cooperative Farming |
| Ecosystem Phenomena | Coevolution, Self-organization, Adaption, and Emergence. |
| Measures | Yield, Productivity, Incomes, Production, Procurement, Acreage, etc, (Refer to Knowledge Graph (iv) in Appendix A) |
| Measures (Ecological) | Soil quality, water quality, air pollution, emissions |

| Metrics (Why) | GDP (Agriculture, per-Capita Farmer), % Export, debt level, poverty level, unemployment level, profitability level, Carbon-Footprint, Energy-related metrics |
|---|---|
| Relator | Banks, Government Agencies, PDS, Insurance Firms, Seed Providers, Fertilizer Firms, etc. |
| Relationships & Networks | Tenancy, Under Organic Farming, Natural Farming, meaningful relationship between any two aspects. |
|  | Farmer Co-operatives, Communes, Trader Associations, Consumer Groups (elaborated further in later part of this section). |
| Separations | Detailed in Table 11 further below. |

## 5.2 Tantra Framework as Normative Framework

In the previous section, we exploited the descriptive power of the Tantra framework. If we want change to happen, we need to make use of the framework in a normative mode. Here we define Tantra Normative Framework in Figure 6 below.

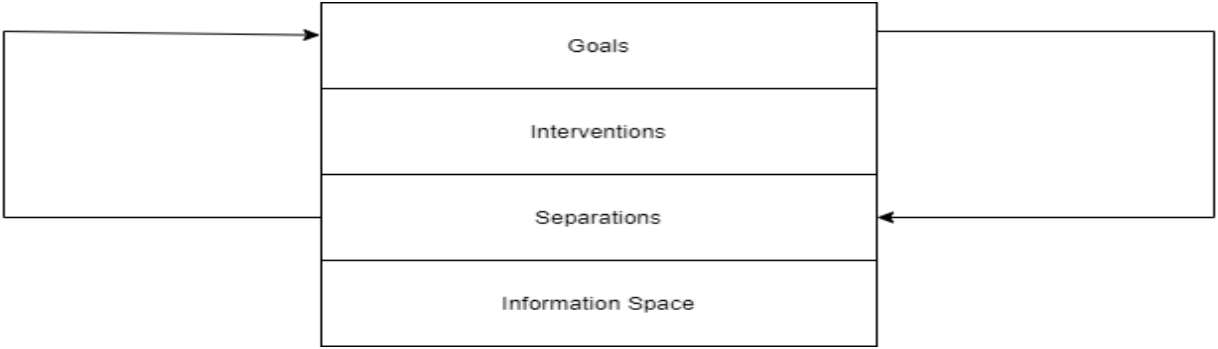

**Figure 6: Tantra Normative Framework**

Here, the process of change comprises of setting goals, designing interventions, and measuring separations (which in turn leads to recalibrating the goals and interventions as things change) operates on information space created using Tantra Framework.

1. We set goals for Agricultural Ecosystem by addressing the subordinate ecosystems namely biological/natural resource, social and economic at production end and business, welfare and industrial at consumption end.
2. To realize goals, governments need to design and deploy interventions. To ensure the right policy interventions are chosen we propose the use of the paradigm of the theory of change [71,72]. Here an intervention is evaluated comprehensively by looking at all assumptions and linkages working backward from outcome to be achieved. The process column of Tantra Framework enables the definition of intervention processes that includes inputs, tasks, outputs, outcomes, and relationships to other processes. Tantra Framework facilitates continued monitoring of interventions.
3. To assess the state of the agricultural sector we use Bartels' theory of market separations. Bartels [73] identified financial, spatial, temporal, and informational separations as inhibitors that come in the way of the healthy functioning of a market. Bartels' theory is particularly suited to the agricultural sector. We add additional interventions that pertain to Agricultural Ecosystem.

The separations are depicted in Table 11 below. The Goals are depicted in Table 12. Table 13 illustrates the analysis of interventions using the Theory of Change Framework.

**Table 11: Market Separations in the Agriculture sector**

| Separation | Remarks |
|---|---|
| **Informational** | Lack of information among farmers on prevailing price bands in the market. |
| **Spatial** | Distant from Mandis/ financial institutions may be a blocker |
| **Temporal** | When a farmer wants to sell, the government may not be buying thus forcing the farmer to sell to aggregators. |
| **Financial** | Inability to access institutional credit at lower interest rates. Inability to reinvest earnings. |
| **Capability** | Ability to shift to crops that have export potential. Ability to absorb risks in case of failure. |
| **Intellectual** | The inability of experts from multiple disciplines to work together |
| **Socio-Political** | Farmers are influenced by their social groups or with whom they transact, making it harder for Government and Experts to influence their perceptions |

**Table 12: Setting Goals for Agricultural Ecosystem**

| Ecosystem | Goals | Remarks |
|---|---|---|
| **Biological Natural Resource Ecosystem** | Improving Soil-Health by cutting down on Fertilizers. Promoting Organic and Natural Farming Promoting Crop Diversification | Goals can be converted to objectives and metrics by using acreage, the volume of crops, and the value of crops |
| **Social Ecosystem** | Encouraging the farmer co-operatives Avoiding any exploitative relationship among Farm-Owners, Tenant Farmers and Farm Workers | Goals can be converted using suitable fairness criteria so that all those who contribute at the production end get what is due to them. |
| **Economic Ecosystem** | Reducing the Fiscal Stress to the Government by optimizing MSP based procurement | Suitable metrics can be designed to optimize the supply chain and value chain. |
| **Business Ecosystem** | Encouraging the use of E-NAM, Sales outside the APMC route | Measurements of the proportion of sales using local Mandi, E-NAM, and sales to private traders, out-of-state traders, and super-markets can be done |
| | Fair distribution of income along the value chain. | We can monitor how income gets distributed among farmers, intermediaries, and traders. |

| **Welfare Ecosystem** | Assessing Farmer Income and Well-being and the difference made by interventions such as PM-KISAN | We can monitor outlays and outcomes related to rural prosperity. |
|---|---|---|
| **Industrial Ecosystem** | Promoting Contract Farming, Leisure, Solar, and Windmill Farming | Goals can be converted to objectives and metrics by using acreage, the volume of crops, and the value of crops. We can assess benefits to consumers. |

Table 13: Theory of Change Methodology applied to Farm Law No. 1(The Farmers Produce Trade and Commerce Bill)

|  | **Farmers selling to any trader outside the APMC system** |
|---|---|
| **Summary Statement** | Giving Freedom to farmers by removing the regulation that they have to sell only in APMC Mandis enabling them to sell to any trader where they do not have to pay any APMC fees. |
| **Problem Statement** | When Farmers are required to sell only at APMC Mandis then they may get into a monopsony situation with traders that have a license. In cases when they sell through commission agents, they may get into monopsony with them, thus reducing the bargaining power. The restriction reduces farmers' profit and raises the prices consumers have to pay. |
| **Overall Goal** | Enabling farmers to sell to whoever they want and whenever they want, thus enabling both spatial and temporal arbitrage |
| **Change Process** | Encouraging farmers to sell directly instead of at Mandis or to the local aggregator or commission agent. |
| **Change Markers** | % of farmers selling outside APMC system, Volume of trade, Value of Trade. Diversity of crops sold directly, Crop-wise change. |
| **Meta-Theory** | When farmers get the freedom to sell without restrictions, they can discover the best possible prices and may be able to sell in less time. Even when they end up selling at Mandis their bargaining power improves because of wider choice. |
| **Inputs** | Appropriate communication to farmers and traders using a variety of media. Spreading the news through communities about the process they can adopt. Platforms/Forums to discover traders and farmers. Mutual Trust/referral/recommendation mechanism. Demand and Supply information. Payment Assurance and Delivery Assurance. |
| **Actors** | Traders, Farmers, Government Departments, Authorities to address any disputes. |
| **Domains of Change** | Agricultural Trading and Commerce |
| **Internal Risks** | Traders accepting goods and not making payments. Farmers accepting payment and not doing timely deliveries. Usual losses in transit. Quality issues. |
| **Assumptions** | The whole change relies on the assumption that farmers are not satisfied with the current APMC system and would like to explore avenues that are not as reliable and well-known as APMC. |

| | | |
|---|---|---|
| **External Risks** | The vested interests may spread misinformation and cause fear, uncertainty, and doubt in farmers. Some trades outside APMC may go bad. |
| **Obstacles to Success** | Farmers are dependent on credit on large farmers and commission agents. This may cause them to continue to live with monopsony. |
| **Knock-On Effects** | Overall efficiency in the market may mean lower prices and better choices to consumers and better returns for farmers. Intermediaries may be forced to add value in terms of logistics, quality assurance, delivery assurance, and payment assurance. Better bargaining power to farmers at APMC and revision of APMC fees to make the Mandis competitive. |

Next, we look at what it takes to operationalize Tantra Framework. Considering the size and complexity of the agricultural sector, we have chosen the ITIL change management methodology of BMC [74], which is typically used in large and complex enterprises. See Table 14 below.

**Table 14**: **Operationalizing the agricultural Ecosystem using Tantra Framework**

| #. | Topic | Remarks |
|---|---|---|
| 1 | **Issues in the current system** | Lack of growth, reduced profitably, stagnant investment in the agricultural sector along with continued poverty and sustained degradation of the environment. |
| 2 | **Change Agents** | Government Departments, Farming communities, and rural populace can act as change agents |
| 3. | **Desired Change** | To promote the agricultural sector as well as the rural economy and spread prosperity among large sections of people. |
| 4 | **List of Services to be built using Tantra Framework** | A digital application/portal needs to be built that makes use of Tantra Framework to capture inputs as well as to share information. |
| 5 | **Strategic Alignment between Objectives and services** | Having high-quality information flow on an ongoing basis can enable sustained social change and progression towards prosperity. |
| 6 | **Design of change Process with Purpose** | Choice of interventions guided by goals and then monitoring them is key. |
| 7 | **Transition Process** | New processes should be instituted covering region after region where data is frequently collected and analyzed. They should interoperate with existing processes harmoniously. |
| 8 | **Organizing actors/Institutions** | Existing government organizations and farmer institutions can be asked to help out while taking care to maintain the sanctity of the data collection process |
| 9 | **Process for collecting feedback** | Feedback from all stakeholders can be collected and shared as appropriate. Feedback should be used to achieve continuous service improvement. |
| 10 | **Training Plan** | The officials and people need to be trained with online material as well as offline programs. |

| 11 | **Future Roadmap** | The scope can be enhanced to include crop analytics in the future. |

## 5.3 Transforming Agricultural Sector from within

Real transformation however can only come from within. Here we elaborate on how farmers can transform the agricultural ecosystem from within. See Table 15 below. Table 16 discusses ecosystem phenomena and the role Tantra Framework can play to analyse them.

**Table 15: Modelling Agriculture Sector as Business Ecosystem**

| **Coevolution** |
|---|
| • Under mutualistic coevolution, Government nudges ecosystem players to seek balanced value capture among participants of the value chain. Consumers should be educated to pay a bit more for food products as a fraction of their consumption basket pay-out. Aggregators, intermediaries, and wholesalers should be nudged to make farming more lucrative by sharing a greater portion of profits with farmers. Again, farmers should give fair pay to farmworkers. Farm owners should pass on benefits to tenant farmers when Government routes the benefits through them. |
| **Self-organization** |
| • Farmers should evolve flexible organizations where they can mutually lease or pool land, labour, capital and produce so that they can collectively profit more.<br>• Traders can participate in E-NAMs and build trusted relationships across the nation.<br>• They can form advocacy groups that balance the interest of farm owners, tenant farmers, farm families, and farmworkers and spread responsible credit culture.<br>• They can engage in cooperative commerce by inter-connecting farming communities and consuming communities directly using digital platforms. There can be buying, selling, transport, delivery, and payment circles that are digitally enabled. |
| **Adaptation** |
| These may be the adoption of ZBNF or organic farming which removes the need for fertilizer subsidies. Choice of drip irrigation to switch over to crops that need less water. From the consumer side choosing Direct Benefit Transfer in place of food grains can be a change. From the standpoint of the Government progressively loosening controls on trading and exports can be the right prescription. |
| **Emergence** |
| Jean-François Ponge[75] proposed 3 different models to explain emergence: bubble(where molecules on the surface give strength to the bubble), crystal(where every part is akin to the whole, and waves model. In the case of the agricultural sector, if stakeholders start to move away from an entitlement mindset towards an empowerment mindset if it spreads like waves, it can radically transform the sector. |

Table 16: Ecosystem Phenomena using Tantra Framework

| Ecosystem Phenomena | Change Markers | Role of Tantra Framework |
|---|---|---|
| **Coevolution** | Cases of Healthy competition among farmers to adopt innovative, /sustainable farming techniques<br>Consumers' willingness to pay more products from sustainable farming,<br>Fair-play among agricultural ecosystem/value chain/supply chain participants<br>Influencing peer farmers to take up contract farming and trade outside APMC | Tantra Framework can capture data and analyze coevolution in the ecosystem using corresponding metrics related to productivity, sustainable practices, reduction in the use of fertilizers, and fair distribution of profits along the value chain including fair sharing of prosperity among farmworkers. Similarly, the extent of peer influence can also be captured using Tantra Framework. |
| **Self-organization** | The extent of farmer participation in Framer cooperatives and the leverage exercised, E-NAM<br>Engaging in Cooperative Commerce by connecting farming communities and consuming communities directly in nearby areas | Tantra Framework can keep track of formal and informal organizations within the ecosystem and the adoption of self-organization as a pervasive phenomenon that makes a significant difference to empower farming, trading, and consuming communities. |
| **Adaptation** | Adapting to reduced dependence on Government and being empowered.<br>Farmers' willingness to participate in crop diversification to prevent ecological degradation.<br>People choosing Flexible Social Benefits/Direct benefits in place of PDS. | Tantra Framework can help analyze how different sections respond to various reforms underway including the newly passed farm laws.<br>The outcome of adaptation by different sections can be quantified and analyzed at the ecosystem level. |
| **Emergence** | Rural prosperity can emerge from an uptick in natural farming, cooperative commerce, contract farming, export-oriented farming, agri-tourism, solar and windmill farms, where farmers are willing to be entrepreneurial. | Tantra Framework can help track emergence in a fine-grained manner. It can also act as a vehicle of communication to detect emergence in pockets and make it widely known thus acting as a force multiplier.<br>. |

## 6. Discussions

The use of ICT for the Agricultural sector is widely established across the world and there are regional, national and global information systems serving the agriculture sector. There are many popular architecture frameworks such as Adaptive Enterprise Architecture, ArchiMate, TOGAF, FAML, ISO/IEC/IEEE 42010, SABSA, and ITIL. However, all of them are focused on different priorities and typically address enterprises and institutions. The 2018-19 Economic Survey of India [76]] did allude to the requirement for "Enterprise Architecture for Governance" where multiple databases are unified. This was followed up with significant initiatives in E-Governance such as IndEA [77] which laid out an enterprise architecture framework for Government at the national level Upon close examination, we concluded that IndEA has an inward and operational focus, making available similar tools to Government that is used by large enterprises. Hence in this work, we have proposed the use of a knowledge-based Tantra Framework that unifies information from multiple ecosystems and addresses ecological, social, economic, financial, and governance-related considerations synergistically.

# 7. Conclusion

The agricultural sector probably provides the most significant public policy challenge to the Government of India. No other sector is as prone to risks and uncertainties resulting in anxieties and apprehensions which test the bounds of rationality. The agricultural sector also provides an unprecedented opportunity to the Government of India, considering the markets, our traditional know-how, research, innovation, and IT capabilities. To make use of the opportunity it is extremely important is to have a transformative vision for the sector and develop new mental maps. We should stop viewing the agricultural sector narrowly in terms of farmer incomes, productivity, and contribution to GDP. Rather, we need to view the agricultural sector as a livelihood, prosperity, and well-being enabler. We need to empower farmers so that they are self-reliant instead of entangling them with all kinds of subsidies and freebies. Despite all the reforms undertaken as well as planned, credit seems to be one area where farmers end up being dependent on individuals and institutions. Evolving a responsible credit culture that all stakeholders respect and serve everyone fairly and expeditiously is the need of the hour. Also, more work is needed to develop a better understanding of the risks and uncertainties inherent in the agricultural sector. Further, we need to understand social psychology that shapes responses towards risks, uncertainties, and change. In the ultimate analysis, as argued by Mehta {78], we need to rekindle the entrepreneurial spirit in Indian farmers. Having an all-encompassing Tantra framework can help achieve all these goals.

# Declarations

## Availability of data and material

Not applicable.

## Code Availability

Not Applicable

## Competing/Conflicting interests

None.

## Funding

None

# Appendix A: Tantra Knowledge Graphs for Agricultural Ecosystem

Here we provide the Knowledge Graphs generated for Agricultural Ecosystem using Tantra Framework.

**(i)** **Persons and Roles Associated with the Agricultural Ecosystem**

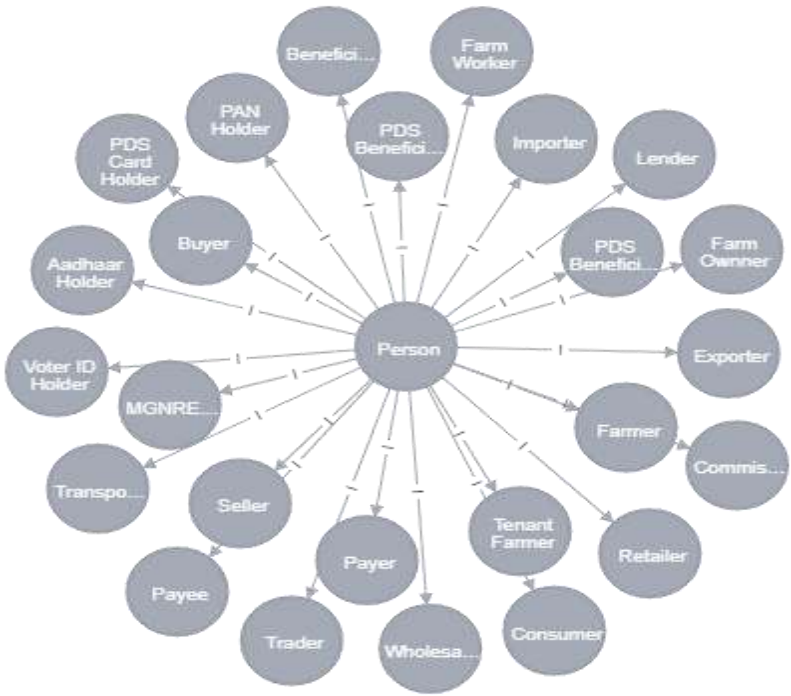

**(ii)** **Types of Farms**

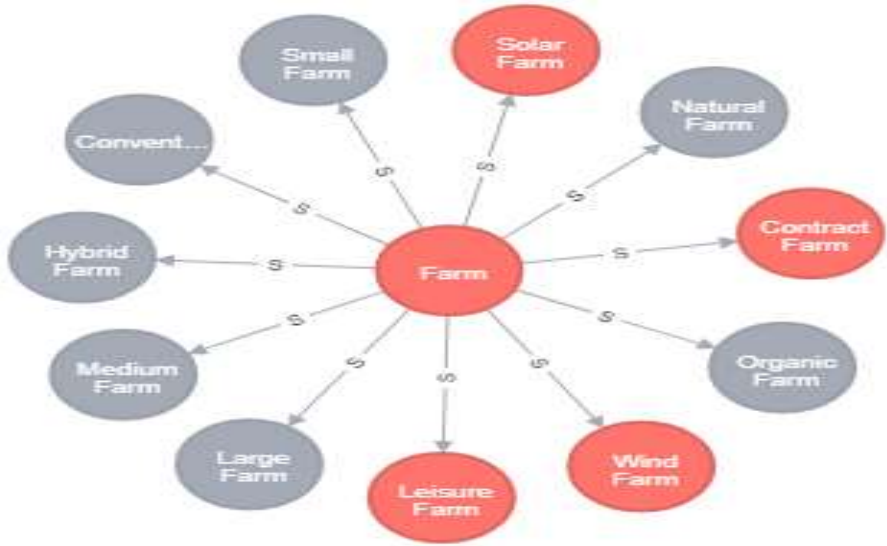

**(iii)  MSP Crops**

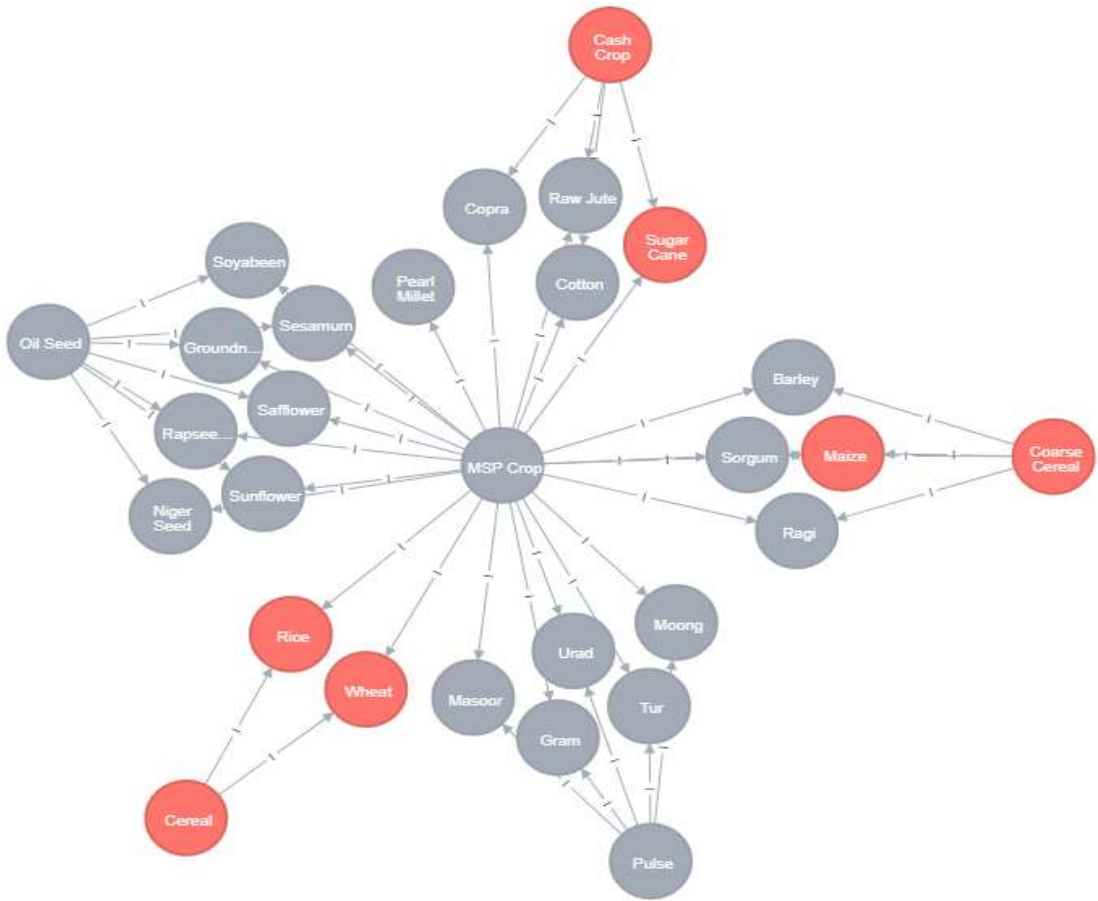

**(iv)  Measures in Agricultural Ecosystem**

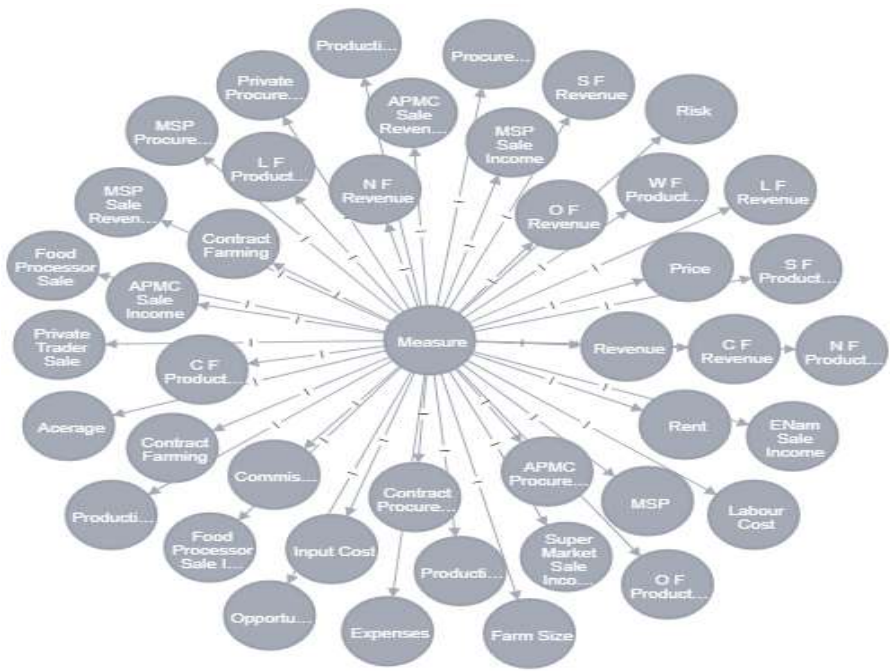